\documentclass{optica-article}
\journal{opticajournal} 
\articletype{Research Article}
\usepackage{lineno}
\usepackage{physics}

\begin{document}

\title{SUPER excitation of quantum emitters is a multi-photon process}

\author{Luca Vannucci\authormark{*} and Niels Gregersen}
\address{DTU Electro, Department of Electrical and Photonics Engineering, 2800 Kongens Lyngby, Denmark}

\email{\authormark{*}lucav@dtu.dk}

\begin{abstract*}

The swing-up of quantum emitter population (SUPER) scheme allows to populate the excited state of a quantum emitter with near-unity fidelity using two red-detuned laser pulses.
Its off-resonant, yet fully coherent nature has attracted significant interest in quantum photonics as a valuable tool for preparing single-photon sources in their excited state  on demand, while simultaneously ensuring straightforward spectral filtering of the laser.
However, the physical understanding of this mechanism in terms of energy exchange between the electromagnetic field and the emitter is still lacking.
Here, we present a fully quantized model of the swing-up excitation and demonstrate that it is in fact a multi-photon process, where one of the modes loses two or more photons while the other gains at least one.
Our findings provide an unexpected physical interpretation of the SUPER scheme and unveil a new non-linear interaction between single emitters and multiple field modes.

\end{abstract*}

\section{Introduction}

Efficient sources of indistinguishable single photons are essential components of optical quantum computers and simulators \cite{Heindel2023, Maring2024}.
Whereas spontaneous parameteric downconversion allows for straightforward generation of highly indistinguishable heralded photons, its probabilistic nature leads to low single-photon purity and collection efficiency, which can only be improved with complex multiplexing schemes \cite{Joshi2018, Kaneda2019}.
On the other hand, quantum emitters based on semiconductor quantum dots yield nowadays highly indistinguishable single photons with high purity and collection efficiency of the order of $\approx 0.6$ \cite{Wang2019a, Tomm2021, Maring2024}.
Activating such sources on demand calls for a fast and coherent optical excitation protocol that prepares the emitter in the desired excited state.

In the so-called swing-up of quantum emitter population (SUPER) scheme \cite{Bracht2021}, two red-detuned laser pulses interacting simultaneously with the emitter cause fast oscillations of the excited state population.
For a proper choice of the laser parameters (frequency, amplitude, and duration of each pulse), the emitter is finally left in the excited state with fidelity close to 100\%.
The technological advantage of this technique, as compared for example with resonant fluorescence, is that laser rejection is straightforwardly achieved via spectral filtering due to the red-detuned nature of the pump \cite{Karli2022, Boos2024}.
Moreover, it populates the excited state directly with a fast and coherent process, as opposed for example to pumping into higher energy levels (which requires incoherent internal relaxation to create the exciton).
This attractive scenario has stimulated intense theoretical and experimental research, including the prospect of applying this technique to cavity-coupled quantum dots \cite{Heinisch2024, Joos2024}, the analysis of the important role played by phonon scattering \cite{Vannucci2023, Vannucci2024, Bracht2022_pssb}, and the possibility of achieving a similar effect with hybrid acousto-optical methods \cite{Kuniej2024}.
While the SUPER scheme was originally devised for semiconductor quantum dots, it relies in fact on general properties of quantum emitters coupled to the electromagnetic field and has now been investigated in different material platforms, such as tin vacancies in diamond \cite{Torun2023} and emitters in the layered transition metal dichalcogenide WSe$_2$ \cite{Vannucci2024}.

In spite of this active interest, the physical origin of this effect is not yet fully understood.
Bracht \textit{et al.} \cite{Bracht2023} have explained the swing-up effect with a dressed state picture of the emitter-field system. While one of the laser pulses dresses the bare emitter states and effectively modifies their energy splitting, the other one drives transition between the dressed states. 
However, a fundamental interpretation in terms of energy exchange between the emitter and the electromagnetic field is still missing.
For example, whereas resonance fluorescence is easily interpreted in terms of absorption of one single photon from the laser pump, such a simple picture is currently unavailable for the SUPER scheme. This is further complicated by the off-resonant nature of this protocol and by the presence of two simultaneous laser fields, which makes the physical interpretation less intuitive.

In this work, we surprisingly show that the SUPER scheme excitation involves multiple photon exchanges between the emitter and the laser. We achieve this by developing a quantum-mechanical model for an emitter coupled to two quantized field modes, which allows to count the number of photons in each mode before and after interaction with the emitter.
Specifically, we find that two photons are typically absorbed from one of the modes during the swing-up process. Of these two excitations, one is stored in the emitter by raising its population from the ground to the excited state, while the remaining one is absorbed into the second field mode.
Interestingly, we find that population inversion may still occur even when one of the field modes is in the vacuum state, provided that the other one is populated with at least two photons.

This article is organized as follows. First, we describe the model and the methodology in Section \ref{sec:model}. Adopting a Fock state assumption for the electromagnetic field, we analyze the dynamics of the SUPER and the red-and-blue dichromatic scheme \cite{Koong2021, Vannucci2023} in the large photon number regime in Sections \ref{sec:SUPER} and \ref{sec:dichromatic}, respectively.
We then focus on the few-photon regime in Section \ref{sec:vacuum}, where we consider the case where one of the field modes is in the vacuum state.
In Section \ref{sec:coherent}, we describe the field in terms of coherent states, and compare the results with the ones obtained under a Fock state assumption.
Then, we touch upon the effect of dissipation and decoherence in Section \ref{sec:decoherence}.
We finally discuss our findings in Section \ref{sec:discussion}, before drawing our conclusions in Section \ref{sec:conclusions}.

\section{Model}
\label{sec:model}

We model the quantum emitter as a two-level system with ground state $\ket{G}$ and excited state $\ket{X}$. Its Hamiltonian is $H_X = \hbar \omega_X \sigma^\dag \sigma$, with $\sigma^\dag = \dyad{X}{G}$ the raising operator. The emitter is coupled to two quantized electromagnetic field modes with Hamiltonian $H_j = \hbar \omega_j a^\dag_j a_j$, $j \in \qty{1, 2}$, where $a_j$ is the annihilation operator for mode $j$.
Using the rotating wave approximation, the emitter-field coupling is modeled as a two-mode Jaynes-Cummings Hamiltonian, 
\begin{equation}
    \label{eq:H_int}
    H_{\rm int}(t) = \sum_{j=1,2} \hbar g_j \exp \qty(-\frac{t^2}{t_p^2}) \qty(\sigma^\dag a_j + \sigma a_j^\dag) .
\end{equation}
The coupling constants $g_j$ in Eq.~\eqref{eq:H_int} are modulated by a time-dependent Gaussian envelope $\exp \qty(-\frac{t^2}{t_p^2})$ to simulate pulsed excitation. The Gaussian modulation ensures a smooth switch-on and switch-off of the interaction and removes additional sidebands from the electromagnetic field spectrum, which would add unwanted features to the system response and hinder the visibility of the SUPER mechanism (see Supplement 1 for details). 

In the single-mode Jaynes-Cummings model, the excitation number $\mathcal N = \sigma^\dag \sigma + a^\dag a$ is a conserved quantity. It follows that the subspace spanned by the pair $\qty{\ket{X, n}, \ket{G, n+1}}$, satisfying $\ev{\mathcal N} = n+1$, is decoupled from the rest of the Hilbert space. This makes it possible to solve the problem analytically, even in the presence of time-dependent coupling \cite{Gruver1993, Joshi1993, Dasgupta1999}.
Similarly, the two-mode Jaynes-Cummings model has a conserved excitation number $\mathcal N = \sigma^\dag \sigma + a_1^\dag a_1 + a_2^\dag a_2$. However, in contrast with the single-mode model, the subspace with constant excitation number $\ev{\mathcal N} = n + 1$ is spanned by the set of states $\qty{\ket{G, n+1, 0}, \ket{X, n, 0}, \ket{G, n, 1}, \ket{X, n-1, 1}, \ldots, \ket{G, 0, n+1}}$ and has dimension $2 (n+1) + 1$, thereby making the analytical solution challenging.

We thus calculate the dynamics numerically by solving the Schrodinger equation $i \hbar \partial_t \ket{\psi(t)} = \widehat H_{\rm int}(t) \ket{\psi(t)}$ in the interaction picture with respect to $H_X + H_1 + H_2$, with 
\begin{align}
    \widehat H_{\rm int}(t)
    & = \exp \qty[\frac{i}{\hbar} (H_X + H_1 + H_2) t] H_{\rm int}(t) \exp \qty[-\frac{i}{\hbar} (H_X + H_1 + H_2) t] \nonumber \\
    & = \sum_{j=1,2} \hbar g_j \exp \qty(-\frac{t^2}{t_p^2}) \qty(e^{-i \delta_j t}\sigma^\dag a_j + e^{+i \delta_j t} \sigma a_j^\dag) ,
\end{align}
and $\delta_j = \omega_j - \omega_X$ the detuning of each mode with respect to the emitter frequency.
Specifically, noting that $\exp \qty(-3^2) \approx 10^{-4}$, we initialize the system in the chosen initial state $\ket{\psi(t_{\rm i})}$ at time $t_{\rm i} = -3 t_p$ and we calculate the evolution of $\ket{\psi(t)}$ with a 4th order Runge-Kutta algorithm until the final state $\ket{\psi(t_{\rm f})}$ at time $t_{\rm f} = +3 t_p$, where the population of the emitter and each field mode is inspected.
Here, the use of the interaction picture is beneficial to avoid fast oscillating terms, especially when dealing with field states with a large number of photons.

It is worth noting that the numerical algorithm does not preserve the state normalization. This is easily verified with a 1st order Euler method by noticing that the evolved state $\ket{\psi(t+\dd{t})} = \ket{\psi(t)} - \frac{i}{\hbar} \dd{t} \widehat H_{\rm int} \ket{\psi(t)}$ satisfies $\braket{\psi(t+\dd{t})} = \braket{\psi(t)} + \frac{(\dd t)^2}{\hbar^2} \braket{\widehat H_{\rm int} \psi(t)} > 1$, even for $\braket{\psi(t)} = 1$. Therefore, we take care of normalizing the quantum state to 1 after each step in the calculation.
This is not strictly necessary to achieve convergence, as the lack of normalization may be compensated with a finer time step $\dd t$. However, it reduces the computation time up to a factor of 10 with respect to the case without state normalization.

For the numerical implementation, the state of each field mode $j$ is expanded on a truncated number state basis $\qty{\ket{n_j}}$, with $n_j \in (N_j^{\rm min}, \ldots, N_j^{\rm max})$, and operators acting on mode $j$ are represented as $M_j \times M_j$ matrices, with $M_j = N_j^{\rm max} - N_j^{\rm min} + 1$.
We make sure that results are converged with respect to the choice of $N_j^{\rm max}$ and $N_j^{\rm min}$. This can be checked, for instance, by ensuring that the expectation value of the excitation number $\mathcal N = \sigma^\dag \sigma + a_1^\dag a_1 + a_2^\dag a_2$ is conserved throughout the simulation, as required by the fact that $\comm{\widehat H_{\rm int}(t)}{\mathcal N} = 0$.

\section{Dynamics of the SUPER scheme}
\label{sec:SUPER}

\begin{figure}
\centering
\includegraphics[width=0.99\linewidth]{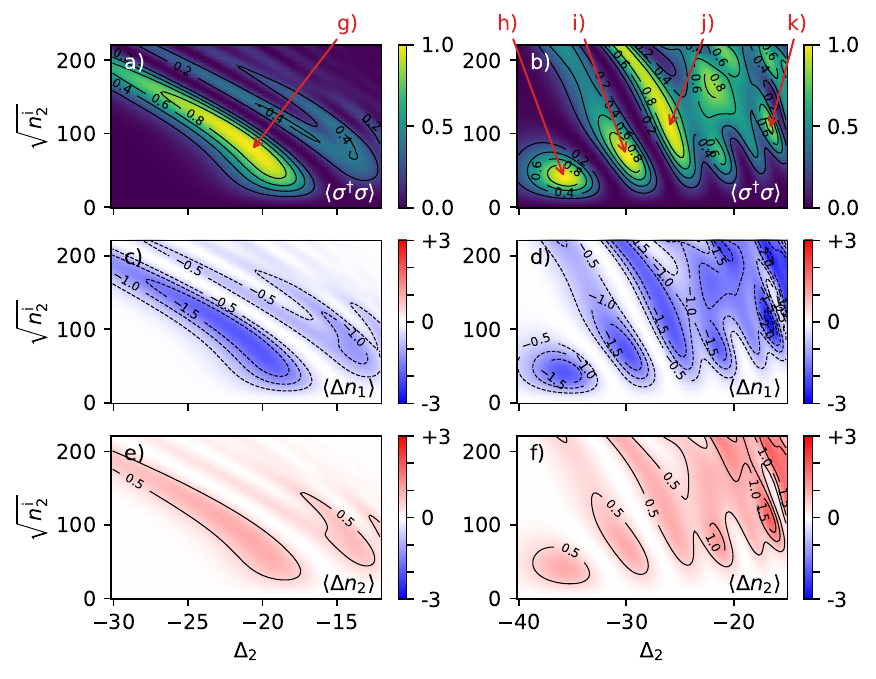}
\includegraphics[width=0.99\linewidth]{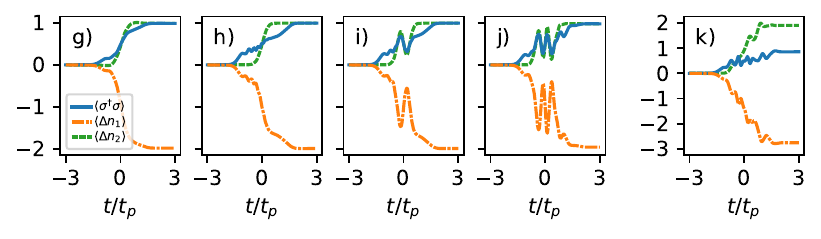}
\caption{
    \textbf{(a--f)}: Quantum state of the coupled emitter-field system after a SUPER pulse, as a function of the detuning $\Delta_2 = t_p \delta_2$ and the initial number of photons $n_2^{\rm i}$ in mode 2.
    The system is initialized at time $t_{\rm i}$ in the state $\ket{G, n_{1}^{\rm i}, n_{2}^{\rm i}}$, with $\ket{n_j^{\rm i}}$ representing Fock states.
    The quantities plotted are the final excited state population $\ev{\sigma^\dag \sigma}$ (panels a, b) and the variation $\ev{\Delta n_j} = \ev{a_j^\dag a_j (t_{\rm f})} - n_j^{\rm i}$ of the average photon number in mode 1 (c, d) and 2 (e, f).
    \textbf{(g--k)}: Evolution of $\ev{\sigma^\dag \sigma}$, $\ev{\Delta n_1}$, and $\ev{\Delta n_2}$ for different configurations corresponding to local maxima in panels a and b, see red arrows.
    Parameters for the first mode are fixed at $\Delta_1 = 6$ (everywhere), $\sqrt{n_1^{\rm i}} = 62.83$ (panels a, c, e), and $\sqrt{n_1^{\rm i}} = 157.08$ (panels b, d, f).
    The dimensionless interaction strength is $G_1 = G_2 = 0.1$ (everywhere).
    }
    \label{fig:1}
\end{figure}

We begin by considering the system dynamics when the field modes are both red-detuned with respect to the emitter and initialized in large-number Fock states $\ket{n_j^{\rm i}}$. 
This is motivated by the fact that optical excitation of a quantum emitter involves laser pulses with a large number of photons.
Although Fock states are generally inaccurate to describe the quantum field of a laser, their use allows for a substantial numerical simplification of the problem in terms of the number $M_j$ of states needed.
The case of coherent states, which represent a more accurate description of the laser field, will be examined later in Section \ref{sec:coherent}, where we find good qualitative agreement between the two pictures.
We note that the dynamics is governed by the dimensionless quantities $\Delta_j = t_p \delta_j$, and $G_j = t_p g_j$, and we will discuss the results in terms of these parameters.

In Fig.~\ref{fig:1}, we explore the population $P_X = \ev{\sigma^\dag \sigma}$ of the $\ket{X}$ state and of both field modes at time $t_{\rm f} = 3 t_p$, i.e.\ after interaction has occurred.
Following the literature \cite{Karli2022, Boos2024, Vannucci2023, Vannucci2024}, we fix the parameters for the first mode (dimensionless detuning $\Delta_1 = t_p \delta_1$ and initial photon number $n_j^{\rm i}$), and scan the corresponding parameters for the second mode.
Note that the terms ``first'' and ``second'' identify the mode with smaller and larger detuning in absolute value, respectively.
In the SUPER scheme literature based on semiclassical models of light-matter interaction, results are typically shown as a function of the electric field amplitude \cite{Bracht2021, Vannucci2023}, which is proportional to the square root of the photon number.
Therefore, we choose to plot the results as a function of $\sqrt{n_2^{\rm i}}$.
The dimensionless coupling is fixed here to $G_1 = G_2 = 0.1$. The case of stronger coupling $G_1 = G_2 = 5$ will be discussed later in Section \ref{sec:vacuum}.
 
In Fig.~\ref{fig:1}a, which is taken at fixed $\sqrt{n_1^{\rm i}} = 62.83$ ($n_1^{\rm i} = 3947$), we identify a region of near-unity exciton preparation fidelity, with a maximum of $P_X = 0.99$.
This is accompanied by additional resonances at smaller detuning with incomplete population inversion of the order of $\approx 0.4$ at maximum.
In contrast, by increasing the initial photon number in mode 1 to $\sqrt{n_1^{\rm i}} = 157.08$ ($n_1^{\rm i} = 24673$), we observe multiple resonances with $P_X \geq 0.99$ (see Fig.~\ref{fig:1}b).
Similar features have been observed in the literature using a classical description of the field and represent the signature of the SUPER scheme. We conclude that our model of a two-level emitter interacting with two quantized field modes captures the physics of the SUPER swing-up effect satisfactorily.

\begin{figure}
\centering
\includegraphics[width=0.99\linewidth]{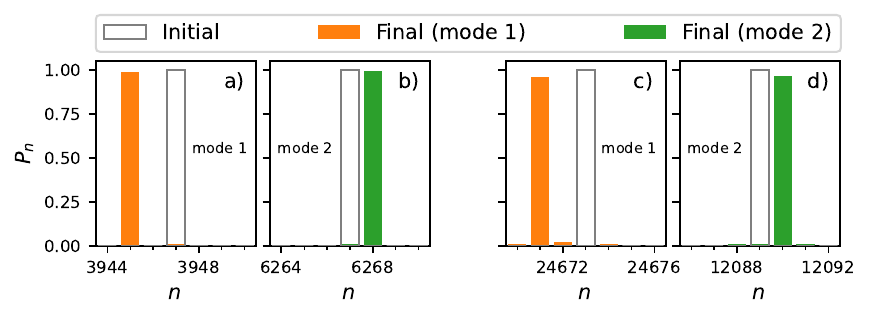}
\caption{
    Photon number distribution of the initial and final state corresponding to Fig.~\ref{fig:1}g (panels a, b) and Fig.~\ref{fig:1}j (panels c, d). Here, the population $P_n$ of each number state $\ket{n}$ is plotted.
    }
    \label{fig:2}
\end{figure}

Differently from the semiclassical description, a quantized model give access to the exact number of photons in each mode. We thus calculate the difference $\ev{\Delta n_j} = \ev{a_j^\dag a_j (t_{\rm f})} - \ev{a_j^\dag a_j (t_{\rm i})}$ of the photon number in each mode before and after interaction, which is plotted in Fig.~\ref{fig:1}c--f.
In correspondence with each resonance of $P_X$, we observe variations of the photon number in both modes. 
Remarkably, mode 1 shows always a decrease in photon number, whereas mode 2 shows an unexpected increase.
Exploring the dynamics in correspondence of $P_X \approx 1$ (see Fig.~\ref{fig:1}g--j), we find that exactly two photons are subtracted from mode 1, while mode 2 gains one photon in the process.
This demonstrates surprisingly that SUPER scheme is a multi-photon process involving a redistribution of photons between the field modes.
We also identify a higher-order process leading to large but incomplete population inversion $P_X > 0.85$ (see Fig.~\ref{fig:1}k) where almost three photons ($\ev{\Delta n_1} = -2.76$) are subtracted from mode 1, while mode 2 gains almost two photons ($\ev{\Delta n_2} = 1.90$).
We hypothesize that a process leading to $\ev{\Delta n_1} = -3$ and $\ev{\Delta n_2} = 2$ exactly might be found with a thorough investigation of the parameter space, which is not performed here.
Despite these variations in the photon number, we observe that the total excitation number $\mathcal N = \sigma^\dag \sigma + a_1^\dag a_1 + a_2^\dag a_2$ is indeed conserved in the process, i.e. $P_X + \ev{\Delta n_1} + \ev{\Delta n_2} = 0$.
The photon number distribution (plotted in Fig.~\ref{fig:2} for two exemplary cases) shows that the final state of modes 1 and 2 is to a very good approximation a Fock state with a well defined photon number.

In contrast with the semiclassical description of the SUPER scheme dynamics, which shows fast and ample oscillations in time of the $\ket{X}$ state population within one pulse (of the order of 10--30), the first leftmost resonance in Fig.~\ref{fig:1}a--b is characterized by a smooth and monotonic dynamics. Higher order resonances at smaller detuning and larger photon number display few oscillations in the emitter population and photon number of the field modes (of the order of 2--3).

\section{Dynamics of the red-and-blue dichromatic scheme}
\label{sec:dichromatic}

In the red-and-blue dichromatic scheme, two laser pulses are symmetrically detuned to the red and blue side of the spectrum with respect to the emitter.
It has been demonstrated that specific configurations involving pulses of different amplitude generate complete population inversion, whereas symmetric configurations with identical pulse amplitude lead necessarily to $P_X=0$ \cite{Koong2021}. Exciton preparation with the red-and-blue scheme occurs through a similar swing-up effect of the population \cite{Koong2021, Vannucci2023}.

\begin{figure}
\centering
\includegraphics[width=0.99\linewidth]{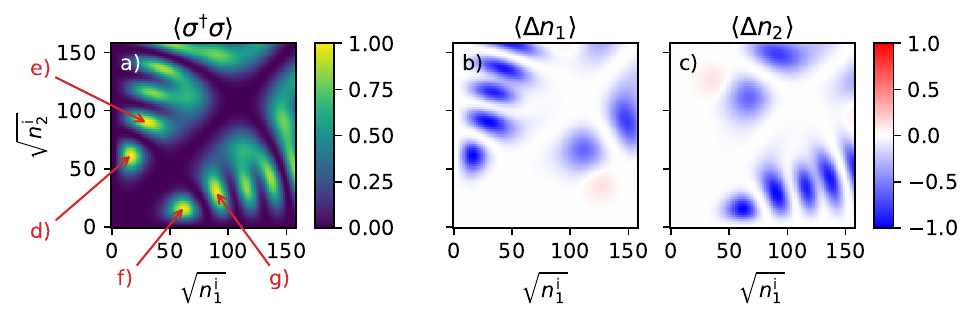}
\includegraphics[width=0.99\linewidth]{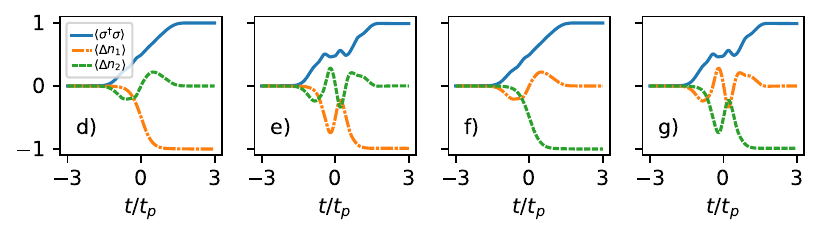}
\caption{
    \textbf{(a, b, c)}: Quantum state of the coupled emitter-field system after a red-and-blue dichromatic pulse, as a function of the initial number of photons $n_j^{\rm i}$ in each mode $j$.
    The system is initialized in the state $\ket{G, n_{1}^{\rm i}, n_{2}^{\rm i}}$, with $\ket{n_j^{\rm i}}$ representing Fock states.
    The quantities plotted are the final $\ket{X}$ state population $\ev{\sigma^\dag \sigma}$ (panel a) and the variations $\ev{\Delta n_j} = \ev{a_j^\dag a_j (t_{\rm f})} - n_j^{\rm i}$ of the average photon numbers in modes 1 (b) and 2 (c).
    \textbf{(d--g)}: Evolution of $\ev{\sigma^\dag \sigma}$, $\ev{\Delta n_1}$, and $\ev{\Delta n_2}$ in time for different configurations corresponding red arrows in panel a.
    In all panels, the mode detunings are fixed at $\Delta_1 = -\Delta_2 = 6$, and the dimensionless interaction strength is $G_1 = G_2 = 0.1$.
    }
    \label{fig:3}
\end{figure}

To model this scenario, we constrain the detuning to $\Delta_1 = - \Delta_2 = 6$ and explore the dependence of the final exciton population as a function of the initial photon number in each mode. Once again, we initialize the field modes in Fock states $\ket{n_j^{\rm i}}$.
As shown in Fig.~\ref{fig:3}a, we obtain several resonances featuring $P_X \approx 1$ in quantitative agreement with results based on a semiclassical model \cite{Vannucci2023}. No resonance is indeed observed along the diagonal $n_1^{\rm i} = n_2^{\rm i}$, in agreement with the requirement of different pulse amplitudes. 

In contrast with the SUPER scheme, we observe that this process involves photon exchange with a single mode, see Figs.~\ref{fig:3}b and Fig.~\ref{fig:3}c.
For example, at $\qty(\sqrt{n_1^{\rm i}}, \sqrt{n_2^{\rm i}})$ = (16.00, 61.25) and (29.21, 89.84), exactly one photon is subtracted from mode 1 (the one at positive detuning) to populate the excited state, while no photon exchange occurs with mode 2 (negative detuning). The opposite holds true by exchanging $n_1^{\rm i}$ with $n_2^{\rm i}$. There, one photon is subtracted from mode 2 with no gain/loss in mode 1.
We notice, however, that in both cases the mode with $\ev{\Delta n_j} = 0$ participates actively in the dynamics, as demonstrated by oscillations of the average photon number in time reported in Fig.~\ref{fig:3}d--g.

Similarly to the SUPER scheme, we notice that the fundamental resonance (namely, the one occurring at smaller photon number) shows a monotonic increase in $P_X$ with no oscillations (Fig.~\ref{fig:3}d--g). Higher order resonances display few small oscillations.

\section{Strong emitter-field coupling: ``vacuum'' swing-up}
\label{sec:vacuum}

So far, the emitter dynamics under two-color excitation with a large number of photons is consistent with a semiclassical approach where the electromagnetic field is treated as a classical object.
We now explore a truly quantum regime where the field modes are populated by a few photons only. 

The strength of the emitter-field coupling is governed by the coupling constants $G_j$ and the electric field amplitude $E_j \propto \sqrt{n_j^{\rm i}}$, as one can readily verify by calculating the matrix elements of $H_{\rm int}$.
Therefore, to maintain the same effective light-matter coupling while scaling down the field occupation to the few photon regime, we simultaneously increase the value of the dimensionless coupling constants $G_j = t_p g_j$.
In experiments, this can be done by increasing the emitter-field coupling $g_j$ via optical engineering of the electromagnetic environment, or by increasing the pulse duration $t_p$.

\begin{figure}
\centering
\includegraphics[width=0.99\linewidth]{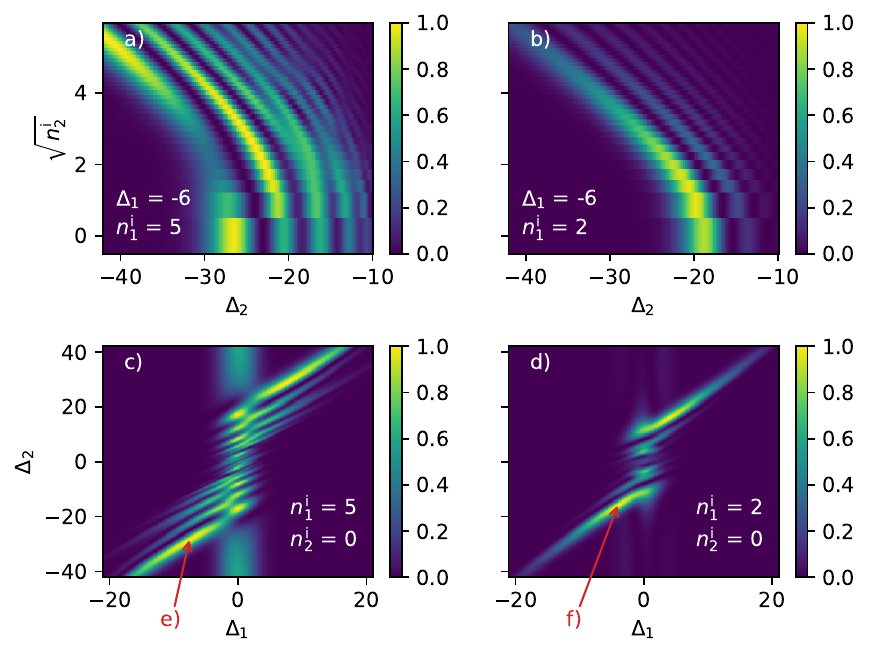}
\includegraphics[width=0.99\linewidth]{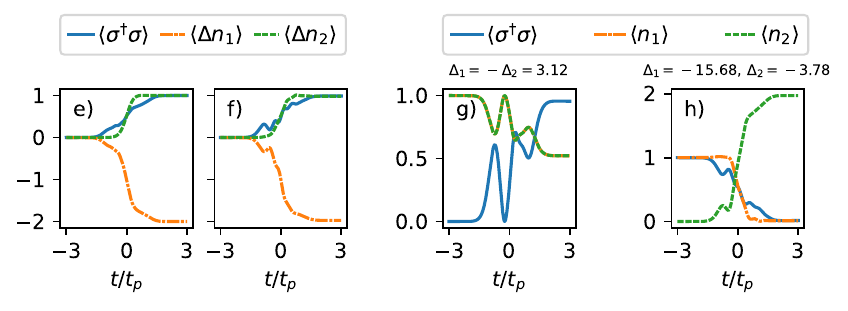}
\caption{
    \textbf{(a--d)}: Final $\ket{X}$ state population $\ev{\sigma^\dag \sigma}$ after a SUPER pulse in the low photon number regime.
    The system is initialized in the state $\ket{G, n_{1}^{\rm i}, n_{2}^{\rm i}}$, with $\ket{n_j^{\rm i}}$ representing Fock states.
    In (a) and (b), the detuning $\Delta_1$ and initial number of photons $n_1^{\rm i}$ in mode 1 are fixed as reported in each panel, $\Delta_2$ is varied continuously along the $x$-axis, and $n_2^{\rm i}$ is varied in integer steps along the $y$-axis.
    In (c) and (d), the initial photon numbers $n_j^{\rm i}$ are fixed and the detunings $\Delta_j$ are varied continuously.
    \textbf{(e, f)}: Evolution of $\ev{\sigma^\dag \sigma}$, $\ev{\Delta n_1}$, and $\ev{\Delta n_2}$ in time for different configurations corresponding to red arrows in panels c) and d).
    \textbf{(g)}: Evolution of $\ev{\sigma^\dag \sigma}$, $\ev{n_1}$, and $\ev{n_2}$ in time for a red-and-blue dichromatic configuration with $\Delta_1 = -\Delta_2 = 3.12$.
    \textbf{(h)}: Evolution of $\ev{\sigma^\dag \sigma}$, $\ev{n_1}$, and $\ev{n_2}$ in time for a SUPER configuration with $\Delta_1 = -15.68$, $\Delta_2 = -3.78$, and where the emiter is initially in the excited state $\ket{X}$.
    Note that panels g) and h) report the absolute photon numbers $\ev{n_j}$ rather than the variation $\ev{\Delta n_j}$.
    In all panels, the dimensionless interaction strength is $G_1 = G_2 = 5$.
    }
    \label{fig:4}
\end{figure}

Starting with the SUPER scheme,  we explore this scenario in Fig.~\ref{fig:4} by increasing the coupling to $G_1 = G_2 = 5$ and scaling the initial number of photons in mode 1 down to $n_1^{\rm i} = 5$ and $n_1^{\rm i} = 2$ (panels a and b, respectively).
Interestingly, we observe that the typical resonances of the SUPER scheme are still clearly visible, and only few photons are needed to achieve full population inversion.
Quite surprisingly, exciton preparation is still possible even when the second field mode is prepared in the vacuum state $n_2^{\rm i} = 0$.
We observe $P_X = 0.97$ for $(n_1^{\rm i}, n_2^{\rm i}) = (5, 0)$ and $P_X = 0.88$ for $(n_1^{\rm i}, n_2^{\rm i}) = (2, 0)$, where incomplete population inversion $P_X < 1$ is explained by the fact that we have chosen an arbitrary value of $\Delta_1 = -6$.
By fixing the initial number of photons in each mode and continuously varying both detunings $\Delta_j$, we obtain indeed full population inversion with $P_X = 1.00$ at $(n_1^{\rm i}, n_2^{\rm i}, \Delta_1, \Delta_2) = (5, 0, -7.56, -28.56)$, and $P_X = 0.99$ at $(n_1^{\rm i}, n_2^{\rm i}, \Delta_1, \Delta_2) = (2, 0, -4.06, -15.96)$, see Figs.~\ref{fig:4}c and \ref{fig:4}d.
The corresponding time-dependent dynamics is plotted in Figs.~\ref{fig:4}e and \ref{fig:4}f, demonstrating that full population inversion with exchange of multiple photons in indeed possible even when one of the field modes is in the vacuum state.
For the case of Fig.~\ref{fig:4}f, where the field is initialized in $(n_1^{\rm i}, n_2^{\rm i}) = (2, 0)$, we notice that mode 1 is left in the vacuum state, while mode 2 is in the single-photon Fock state $\ket{n_2^{\rm f}} = \ket{1}$ after interaction.
It should be noted that the Gaussian modulation of the interaction plays a minor role in the case of a vacuum state, i.e.\ we can still explain the vacuum swing-up qualitatively if we remove the $\exp(-t^2/t_p^2)$ factor from $g_2$ (see Supplement 1 for details).

We observe no population inversion when $(n_1^{\rm i}, n_2^{\rm i}) = (1, 0)$, for any value of the detunings.
This demonstrates that the minimum excitation number to activate the SUPER mechanism is $\mathcal N = 2$, consistent with the fact that the SUPER scheme cannot be explained with the exchange of one single photon between the emitter and the field.

Applying similar considerations, we find signatures of the red-and-blue dichromatic scheme in the few-photon regime as well.
Despite the fact that dichromatic excitation in the large photon number regime is explained with the subtraction of a single photon from one of the modes, we do not observe population inversion for initial state $(n_1^{\rm i}, n_2^{\rm i}) = (1, 0)$ or $(0, 1)$, indicating that the requirement on the excitation number is still $\mathcal N \geq 2$.
On the other hand, we find that preparing the field modes in two single-photon states $\qty(n_1^{\rm i}, n_2^{\rm i}) = (1, 1)$ at detuning $\Delta_1 = -\Delta_2 = 3.12$ leads to population of the exciton state with fidelity $P_X = 0.96$, see Fig.~\ref{fig:4}g.
Here, the system is approximately left in the entangled state $\ket{\psi(t_{\rm f})} = \frac{1}{\sqrt{2}} \qty(\ket{X, 1, 0} + \ket{X, 0, 1})$ after interaction, meaning that half a photon has been subtracted from both modes on average.

Finally, it is worth noting that the multi-photon resonances of the SUPER scheme work in the reverse fashion as well. In other words, it is possible to completely depopulate an initially excited emitter by subtracting one excitation from mode 1, while increasing the population of the other field mode by two excitations.
This may be used to engineer non-linear interaction between field modes at different frequency. For example, as shown in Fig.~\ref{fig:4}h, a single-photon state at detuning $\Delta_1 = -15.68$ is converted into a two-photon state at $\Delta_2 = -3.78$ accompanied by relaxation of the emitter to the ground state.

%
%
%
%
%

\section{Coherent states}
\label{sec:coherent}

\begin{figure}
\centering
\includegraphics[width=0.99\linewidth]{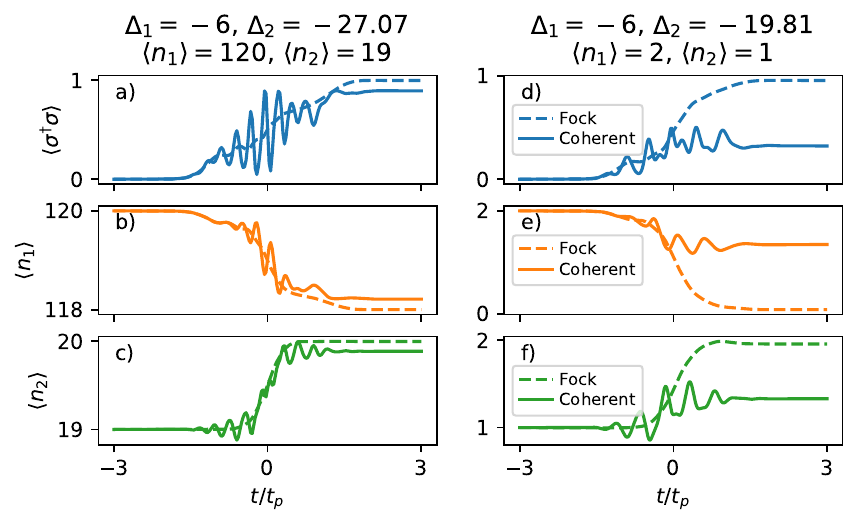}
\includegraphics[width=0.99\linewidth]{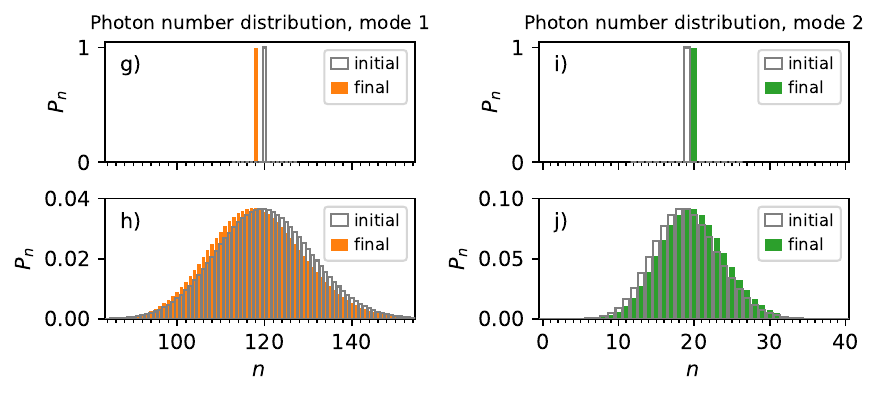}
\caption{
    \textbf{(a--f)}: Evolution of $\ev{\sigma^\dag \sigma}$, $\ev{n_1}$, and $\ev{n_2}$ in time for different initial states of the electromagnetic field.
    The field modes are initialized either in Fock states containing $n_1^{\rm i}$ and $n_2^{\rm i}$ photons respectively (dashed lines), or in coherent states with the same average number of photons (solid lines).
    Detunings $\Delta_j$ and initial photon numbers $n_j^{\rm i}$ are indicated in the top panel title.
    The dimensionless interaction strength is $G_1 = G_2 = 1$ (a--c) and $G_1 = G_2 = 5$ (d--f).
    \textbf{(g, h)}: Initial and final photon number distribution in mode 1 corresponding to panel b for Fock and coherent states, respectively.
    \textbf{(i, j)}: Initial and final photon number distribution in mode 2 corresponding to panel c for Fock and coherent states, respectively.
    }
    \label{fig:5}
\end{figure}

So far we have considered Fock states of the field modes for simplicity.
However, the quantum state of a laser pulse is best described by a coherent state $\ket{\alpha} = e^{-|\alpha^2|/2} \sum_n \alpha^n / (\sqrt{n!}) \ket{n}$. 
In Fig.~\ref{fig:5}, we compare the dynamics for a case where the field is initialized in the multi-mode Fock state $\ket{n_1^{\rm i}, n_2^{\rm i}}$ with the case where the initial field state is $\ket{\alpha_1^{\rm i}, \alpha_2^{\rm i}}$, with $\ket{\alpha_j^{\rm i}}$ representing coherent states containing the same number of photons $\mel{\alpha_j^{\rm i}}{a_j^{\dag} a_j}{\alpha_j^{\rm i}} = |\alpha_j^{\rm i}|^2 = n_j^{\rm i}$ on average.
For sufficiently high excitation number (see Fig.~\ref{fig:5}a--c), we observe that the two situations are qualitatively similar. Indeed, the description based on coherent states confirms that the SUPER scheme excitation takes place by removing approximately two photons from mode 1 while simultaneously increasing the population of mode 2 by one photon.
However, whereas the Fock state description yields a monotonic dynamics, the coherent state model reproduces the typical swing-up pattern that characterizes the SUPER scheme, resulting in fast oscillations of the exciton population.
Despite this remarkable qualitative difference, the attained value of $P_X$ using coherent states is quantitatively similar to the one obtained using Fock states.
An inspection of the photon number distribution, see Fig.~\ref{fig:5}g--j, confirms the agreement between the two pictures. In both cases, the distribution of mode 1 (2) is rigidly shifted towards smaller (larger) values by the same amount. This also shows that both Fock and coherent states maintain their characteristics in terms of number distribution.

However, the agreement breaks down in the deep quantum regime of low excitation number. As shown in Figs.~\ref{fig:5}d--f, a model using coherent states predicts $P_X = 0.32$ for $(|\alpha_1^{\rm i}|^2, |\alpha_2^{\rm i}|^2) = (2, 1)$ and $(\Delta_1, \Delta_2) = (-6, -19.81)$, while the value predicted with a Fock state assumption is $P_X = 0.96$.
We conclude that SUPER scheme excitation in the low photon number regime necessitates of few-photon Fock states as compared with coherent states of small amplitude. Whereas the latter can be realized, for example, with attenuated laser pulses, the former are fundamentally different and more challenging to realize \cite{Cooper2013, Uria2020, Sonoyama2024}.


\section{Dissipation and decoherence}
\label{sec:decoherence}

Before concluding, we briefly examine the effect of dissipation and decoherence on the multi-photon scattering process.
We limit the discussion to (i) spontaneous decay and (ii) pure dephasing of the two-level emitter.
A more detailed modeling of noise and decoherence, including for example phonon scattering beyond the pure dephasing approximation, lies outside the scope of this work.

In both cases, the dynamics becomes non-unitary and an open quantum system approach is necessary.
We thus solve the master equation for the system density operator $\rho(t)$,
\begin{equation}
    \dv{\rho}{t} = -\frac{i}{\hbar} \comm{\widehat H_{\rm int}(t)}{\rho} + \Gamma \mathcal L_\sigma [\rho] + \eta \mathcal L_{\sigma^\dag \sigma} [\rho]
\end{equation}
where spontaneous decay at rate $\Gamma$ and pure dephasing at rate $\eta$ are included, with 
\begin{equation}
    \mathcal L_A[\rho] = A \rho A^\dag - \frac{1}{2} \acomm{A^\dag A}{\rho} .
\end{equation}
Here, the density operator is represented as a $2 M_1 M_2 \times 2 M_1 M_2$ matrix on the same truncated number state basis $\qty{\ket{n_j}}$ as before.

\begin{figure}
\centering
\includegraphics[width=0.99\linewidth]{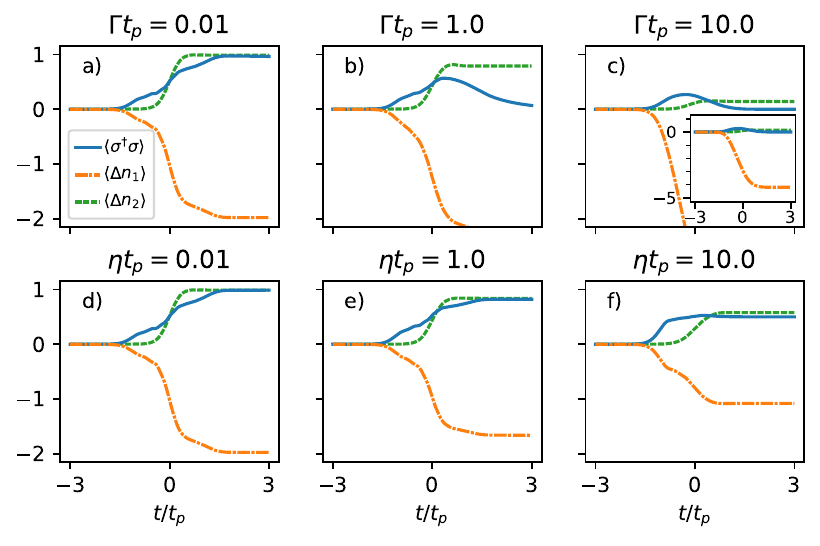}
\caption{
    Effect of \textbf{(a--c)} spontaneous emission at rate $\Gamma$ and \textbf{(d--f)} pure dephasing at rate $\eta$ on the multi-photon scattering.
    The pulse duration is $t_p = 1$ ps. Other parameters are as in Fig.~\ref{fig:4}e.
    }
    \label{fig:6}
\end{figure}

In Fig.~\ref{fig:6}a--c, we consider the configuration previously shown in Fig.~\ref{fig:4}e (i.e. $n_1^{\rm i} = 5$ and $n_2^{\rm i} = 0$, with $G_1 = G_2 = 5$) and we add spontaneous emission with increasing rate $\Gamma$. Here, we fix the pulse duration to $t_p = 1$ ps and calculate the corresponding detunings $\delta_j$ and coupling strengths $g_j$ to obtain the same values of $\Delta_j = t_p \delta_j$ and $G_j = t_p g_j$ as in Fig.~\ref{fig:4}e.
We observe that the SUPER resonance is progressively damped with increasing $\Gamma$. 
Specifically, the dynamics is unaffected when $\Gamma t_p = 0.01$, i.e. when the cavity-emitter interaction occurs on a timescale $t_p$ that is much smaller than the exciton lifetime $1/\Gamma$.
On the other hand, no population inversion occurs in the opposite limit of $\Gamma t_p = 10$, where the exciton lifetime is smaller than the interaction time and dissipation dominates. Here, we observe that the cavity population is depleted almost entirely, see the inset of Fig~\ref{fig:6}c.

We observe similar features when considering pure dephasing at rate $\eta$, see Fig.~\ref{fig:6}d--f. The dynamics is almost unaffected if $\eta t_p \ll 1$, whereas the multi-photon scattering is hindered for larger values of $\eta t_p$ (although it is still partially visible at $\eta t_p = 10$).

\section{Discussion}
\label{sec:discussion}

We have investigated the physics of two-color excitation schemes in terms of energy exchanges between the emitter and the laser pump.
To do so, we have developed a model for a two-level emitter interacting with two quantized field modes, where interaction is modulated with Gaussian pulses of duration $t_p$ in time.
Both the SUPER and the red-and-blue dichromatic schemes have been considered.

We have observed that exciton population under the SUPER scheme, where both modes are red-detuned with respect to the emitter, is accompanied by variations in the photon number of both modes, unveiling the multi-photon nature of this process.
We find clear signatures of events with $(\ev{\sigma^\dag \sigma}, \ev{\Delta n_1}, \ev{\Delta n_2}) = (+1, -2, +1)$ and $(+1, -3, +2)$, where multiple excitations are subtracted from the 1st field mode (the one with smaller detuning, in absolute value). While one excitation is used to raise the emitter population to the excited state, the remaining are absorbed by the 2nd field mode at larger detuning.
Interestingly, this holds true in the regime where the field modes are populated with only few photons, and even when one of the field is in the vacuum state.
This last scenario requires precisely tailored nanophotonic environment that supports the two modes of interest at the required detuning, to avoid accidental interaction with other unoccupied modes.
We speculate that events involving exchange of higher number of photons, i.e.\ $(\ev{\sigma^\dag \sigma}, \ev{\Delta n_1}, \ev{\Delta n_2}) = (1, -m, m-1)$ with $m>3$, might be found at larger total power, although we have not observed such a case in this work.

Interestingly, we find that energy conservation is violated by the multi-photon scattering event.
Before interaction, considering an initial state $\ket{G, n_1^{\rm i}, n_2^{\rm i}}$, the total energy is
\begin{equation}
	E_{\rm i} = \hbar \qty(\omega_X + \delta_1) n_1^{\rm i} + \hbar \qty(\omega_X + \delta_2) n_2^{\rm i}
\end{equation}
Following a variation of $\ev{\Delta n_j}$ in the average photon numbers, the final energy is
\begin{equation}
	E_{\rm f} = \hbar \qty(\omega_X + \delta_1) \qty(n_1^{\rm i} + \ev{\Delta n_1}) + \hbar \qty(\omega_X + \delta_2) \qty(n_2^{\rm i} + \ev{\Delta n_1}) + \hbar \omega_X \ev{\sigma^\dag \sigma}
\end{equation}
It follows that the difference $\Delta E = E_{\rm f} - E_{\rm i}$ is 
\begin{equation}
	\Delta E = \ev{\Delta n_1} \hbar \delta_1 + \ev{\Delta n_2} \hbar \delta_2 + \hbar \omega_X \qty(\ev{\Delta n_1} + \ev{\Delta n_2} + \ev{\sigma^\dag \sigma}) = \ev{\Delta n_1} \hbar \delta_1 + \ev{\Delta n_2} \hbar \delta_2,
\end{equation}
independently of $\omega_X$, with $\hbar \omega_X \qty(\ev{\Delta n_1} + \ev{\Delta n_2} + \ev{\sigma^\dag \sigma})$ vanishing because the excitation number $\mathcal N$ is conserved.
For the fundamental resonance $(\ev{\sigma^\dag \sigma}, \ev{\Delta n_1}, \ev{\Delta n_2}) = (+1, -2, +1)$, we thus have $\Delta E = \hbar \qty(\delta_2 - 2 \delta_1)$. 
In our work, we always find that $|\delta_2| > 2 |\delta_1|$ is needed to obtain population inversion close to 1, consistently with the original prediction formulated within a classical approximation for the electromagnetic field \cite{Bracht2021, Bracht2023}. It follows necessarily that $\Delta E = \hbar \qty(\delta_2 - 2 \delta_1) \ne 0$.
This violation might be explained by the time-dependent nature of the Hamiltonian. Indeed, energy conservation requires an Hamiltonian that is translational invariant in time, in the same way as momentum conservation requires a translational invariant Hamiltonian in space.

This finding is in stark contrast with the paradigm of resonant excitation, where the emitter exchanges one energy quantum with a single field mode and energy is overall conserved.
We have indeed verified that our model reproduces this scenario by removing the 2nd field mode (i.e.\ setting $G_2=0$) and moving the 1st mode into resonance with the emitter ($\Delta_1 = 0$).
By scanning the initial number of photons $n_1^{\rm i}$ in mode 1, we have found clear signatures of Rabi oscillations. For configurations resulting in $P_X = 1$, we have observed that one single photon is removed from the field.
A similar scenario is expected to take place in phonon-assisted excitation \cite{Cosacchi2019, Gustin2020, Thomas2021}, although we have not included phonon coupling in this work.

Reversing the working principle of the SUPER scheme, our model shows that it is possible to stimulate emission from an excited emitter with a simultaneous redistribution of photons among the field modes.
For instance, a system initialized in the state $\ket{X, n_1, n_2}$  is found in the state $\ket{G, n_1 - 1, n_2 + 2}$ after interaction with two red-detuned modes. Once again, this is valid in the case $n_2=0$ where mode 2 is in the vacuum state.
Quite interestingly, this observation challenges the typical paradigm of spontaneous emission stating that the emitted photon must have the same frequency as the photons in the incident wave.
These findings may be used to engineer a non-linear interaction where one single photon stimulates the emission of two identical photons at a different frequency.

Considering the red-and-blue dichromatic excitation, where one of the field modes is moved to the blue side of the spectrum, we have observed that it involves energy exchange with only one of the field modes.
However, our results demonstrate that the presence of a second field mode is necessary to engineer full population inversion, even when no variation in its the photon number is recorded.
An exception to this paradigm is the case of dichromatic excitation for an initial state $\ket{G, 1, 1}$, where we find that half a photon is subtracted on average from both modes.

It is worth noting that we have obtained a good qualitative agreement with the description of the SUPER and red-and-blue dichromatic scheme under a semiclassical model for light-matter interaction, where the field is not quantized (see also Supplement 1).
However, the advantage of the quantum model presented here is that it makes it possible to calculate the variation in photon number for each mode.
We have furthermore observed that a description of the laser field in terms of coherent states is needed to reproduce the typical swing-up behavior of the excited state population under SUPER scheme excitation, which is characterized by fast oscillations.
However, we have surprisingly found that a description in term of Fock states yields qualitatively similar results for the final population inversion, and demands less computational resources.
For example, the Fock state dynamics shown in Fig.~\ref{fig:5}a is fully converged when using $M_1 = M_2 = 11$ states in the field Hilbert space, whereas $M_1 = M_2 = 91$ states are required for full convergence of the corresponding coherent state simulation.
The qualitative agreement between the two descriptions is however limited to the large photon number regime.
For small photon numbers of the order of few units, fundamental differences in the behavior of Fock and coherent states emerge.

Finally, we have used values in the range 0.1--5 for the dimensionless coupling constants $G_j = g_j t_p$. 
Considering short pulses of the order of $t_p =$\,2--3 ps, a value $G_j = 0.1$ is obtained with a light-matter coupling strength of the order of $g_j =$\,0.03--0.05 THz. The latter corresponds to typical values attained in high-Q micropillars \cite{Wang2020_BiYing, Vannucci2023} or open tunable microcavities \cite{Tomm2021}, and similar values are found elsewhere in the literature \cite{Gustin2020}.
Two strategies may be pursued to attain stronger dimensionless coupling $G_j = 5$. On one hand, the pulse duration may be increased by one order of magnitude to $t_p \approx 100$ ps. We notice, however, that the system in this regime is more prone to environment induced decoherence, as proven in Fig.~\ref{fig:6}.
As alternative, a light-matter coupling strength $g_j \approx 2$ THz is needed to obtain $G_j = 5$.
Whereas such a strong coupling maybe challenging to realize, we notice that nanocavities with extreme photon confinement well below the diffraction limit in dielectrics have been designed and fabricated in silicon \cite{Albrechtsen2022}.
Here, ultra-strong light-matter coupling is expected due to the extremely low mode volume.

During the writing of this manuscript, we became aware of a related work which also demonstrates the multi-photon nature of the SUPER scheme \cite{Richter2024}.

\section{Conclusions}
\label{sec:conclusions}

We have studied the exchange of energy between a two-level emitter and two quantized field modes under the SUPER scheme excitation.
Surprisingly, we have found that one mode loses two or more photons while the other gains at least one, proving that the SUPER is a multi-photon process involving a redistribution of photons between the field modes.
Our results unveil a novel and unexpected off-resonant interaction of light and matter, and open new possibilities for manipulating the quantum state of atoms using only few off-resonant photons.

\begin{backmatter}

\bmsection{Funding}
This work received supported from the European Research Council (ERC-CoG ``Unity'', grant no.865230) and the European Union's Horizon 2020 Research and Innovation Programme under the Marie Skłodowska-Curie Grant (Agreement No. 861097).


\bmsection{Disclosures}
The authors declare no conflicts of interest.

\bmsection{Data availability} Data underlying the results presented in this paper may be obtained from the authors upon reasonable request.

\bmsection{Supplemental document} See Supplement 1 for supporting content.

\end{backmatter}

\bibliography{biblio}

\end{document}


\maketitle

\section{Role of the time-dependent Gaussian envelope of the coupling constants}

In this Supplementary Note, we investigate the role of the time-dependent Gaussian envelope $\exp \qty(-t^2 / t_p^2)$ of the coupling constants in Eq.~(1) of the main text.

Light-matter interaction between a single two-level emitter and a quantized cavity mode is typically modeled with a time-independent Jaynes-Cummings Hamiltonian.
In a frame rotating at the bare exciton frequency $\omega_X$, and considering two quantized modes of the electromagnetic field, the Hamiltonian reads
\begin{align}
    H = \sum_{j=1}^{2} \hbar \delta_j a_j^\dag a_j + \sum_{j=1}^{2} \hbar g_j \qty(\sigma^\dag a_j + \sigma a_j^\dag)
\end{align}
where $\sigma^\dag$ raises the emitter population from the ground ($\ket{G}$) to the excited ($\ket{X}$) state, $a_j^\dag$ creates one photon with energy $\hbar (\omega_X + \delta_j)$ in mode $j$, and we make use of the rotating wave approximation.

In Eq.~(1) of the main text, we perform the substitution
\begin{equation}
    g_j \to g_j \exp \qty(-\frac{t^2}{t_p^2})
\end{equation}
to introduce a time-dependent Gaussian modulation of the coupling constants, with identical shape and length $t_p$ of the pulse for each mode.
The Gaussian envelope is chosen to simulate the scenario of pulsed laser excitation, as typically done when the electromagnetic field is treated as a classical (i.e. non-quantized) object \cite{Bracht2021, Vannucci2023, Gustin2020}.
Moreover, the smooth envelope avoids an abrupt switch-on and switch-off of the interaction, which introduces additional features in the system response as we demonstrate here below.

\begin{figure}
\centering
\includegraphics[width=.99\linewidth]{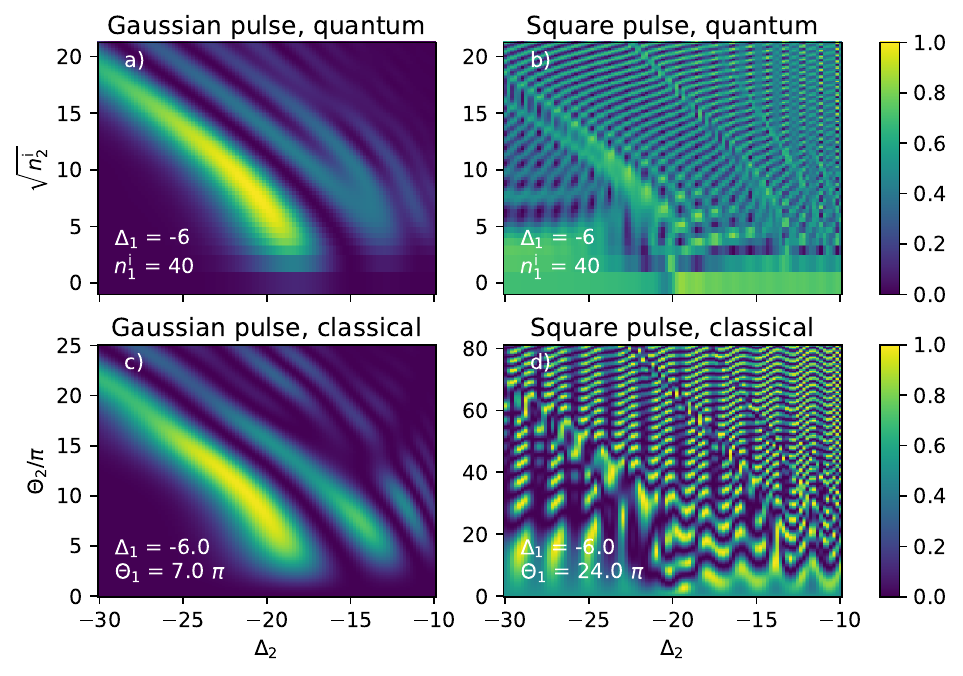}
\includegraphics[width=.99\linewidth]{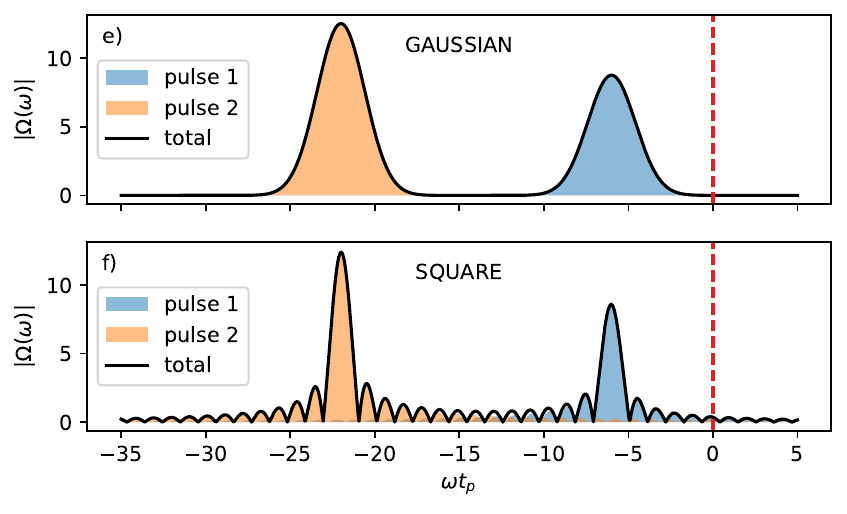}
\caption{
    \textbf(a--d): Final population of the $\ket{X}$ state after a SUPER pulse predicted by a quantized (a, b) and a classical (c, d) model of the field.
    Parameters for the first pulse are fixed as reported in each panel.
    \textbf(e, f): Frequency spectrum for Gaussian (e) and Square (f) pulses in a classical model. Parameters are $\Theta_1 = 7\pi$, $\Theta_2 = 10\pi$, $\Delta_1 = -6$, $\Delta_2 = -22$, corresponding roughly to the maximum in panel c. 
}
\label{fig:S1}
\end{figure}

In Fig.~\ref{fig:S1}a and Fig.~\ref{fig:S1}b, we plot the final excited state population obtained with and without the time-dependent envelope, respectively.
Apart from removing the time-dependent modulation, all parameters in Fig.~\ref{fig:S1}b are identical to Fig.~\ref{fig:S1}a.
It should be noted that the removal of $\exp \qty(-t^2/t_p^2)$ has the same effect as replacing the Gaussian pulse with a Square pulse of duration $6 t_p$, since we run calculations from the initial time $t_{\rm i} = -3 t_p$ to $t_{\rm f} = 3 t_p$.
We observe that the typical SUPER resonance of Fig.~\ref{fig:S1}a is replaced by a rich landscape in Fig.~\ref{fig:S1}b characterized by fast oscillations, where it is difficult to identify a clear resonance.

It is instructive to compare this scenario to the results obtained under a classical approximation of the electromagnetic field.
We thus introduce the semi-classical Hamiltonian
\begin{align}
    H_{\rm sc} = \frac{\hbar}{2} \sum_{j=1}^2 \Omega_j(t) \qty[ e^{-i \delta_j t} \sigma^\dag + e^{i \delta_j t} \sigma]
\end{align}
where the envelope $\Omega_j(t)$ is
\begin{equation}
\label{eq:Gauss}
    \Omega_{j,{\rm Gauss}}(t) = \frac{\Theta_j}{\sqrt{\pi} t_p} \exp \qty(-\frac{t^2}{t_p^2})
\end{equation}
for a Gaussian pulse and 
\begin{equation}
\label{eq:Square}
    \Omega_{j,{\rm Square}}(\omega) = \frac{\Theta}{6 t_p} \qty[\mathrm{Heav}\qty(t + 3 t_p) - \mathrm{Heav}\qty(t - 3 t_p) ] .
\end{equation}
for a Square pulse of length $6t_p$.
Here, $\Theta_j = \int_{-\infty}^{+\infty} \dd{t} \Omega_j(t)$ is the pulse area, and $\mathrm{Heav}(x)$ is the Heaviside step function.
We report the results under a semiclassical model in Figs.~\ref{fig:S1}c and \ref{fig:S1}d, where we observe excellent agreement between the two models. Importantly, the classical model shows the same oscillating pattern when using Square pulses, with no clear signature of the SUPER scheme.

To understand the origin of these fast oscillations, we consider the frequency content of each pulse shape. The Fourier transforms of \eqref{eq:Gauss} and \eqref{eq:Square} read
\begin{align}
\label{eq:Gauss_F}
    \Omega_{j,{\rm Gauss}}(\omega) & = \frac{\Theta_j}{\sqrt{2 \pi}} \exp \qty[-\frac{t_p^2}{4}(\omega - \delta_j)^2 ] , \\
\label{eq:Square_F}
    \Omega_{j,{\rm Square}}(\omega) & = \frac{\Theta_j}{\sqrt{2 \pi}} \frac{\sin \qty[3 t_p (\omega - \delta_j)]}{3 t_p (\omega - \delta_j)}.
\end{align}
Their remarkable qualitative difference is plotted in Figs.~\ref{fig:S1}e and \ref{fig:S1}f, where we show the total pulse $\Omega(\omega) = \Omega_1(\omega) + \Omega_2(\omega)$ together with the individual contributions.
The Gaussian profile remains Gaussian in the frequency domain and is well separated from the exciton frequency ($\omega=0$ in this frame) if the detunings are sufficiently large.
On the other hand, the spectrum of a Square pulse has multiple satellite peaks which may overlap with the exciton, leading to a rich and complicated system response.
This justifies our choice of including a time-dependent modulation of the coupling constants in the Jaynes Cummings Hamiltonian to model the quantized version of the SUPER scheme.

\begin{figure}
\centering
\includegraphics[width=.99\linewidth]{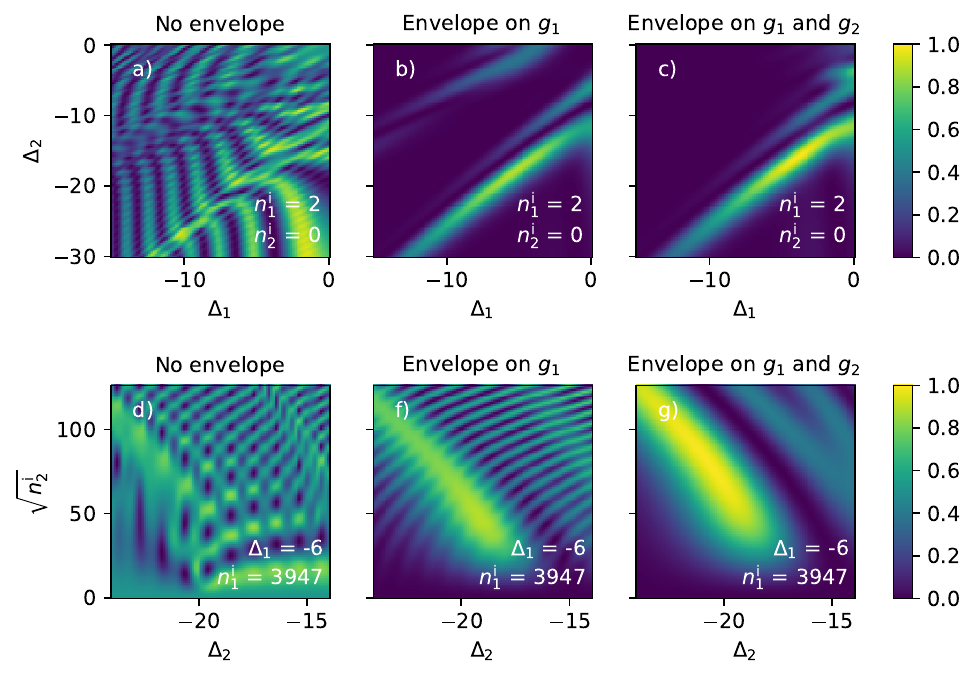}
\caption{
    Final $\ket{X}$ state population $\ev{\sigma^\dag \sigma}$ predicted with a quantized model of the field, with and without a Gaussian envelope of the coupling constants $g_j \exp \qty(-\frac{t^2}{t_p^2})$. 
    In (a--c), a clipping of Fig.~4d of the main text is shown. 
    In (d--e), a clipping of Fig.~1a of the main text is shown.
    }
\label{fig:S2}
\end{figure}

Finally, it is interesting to explore the effect of the Gaussian envelope on the coupling constants $g_1$ and $g_2$ separately, especially when dealing with the ``vacuum'' swing-up effect described in Section 5 of the main text.
We thus consider in Fig.~\ref{fig:S2} a third hybrid case where $g_1$ is modulated with a time-dependent Gaussian profile, while $g_2$ is not.
We observe that the modulation of $g_1$ alone is sufficient to explain qualitatively the results shown in Fig.~4d of the main text, where mode 2 is initially in the vacuum state.
On the contrary, a modulation of both $g_1$ and $g_2$ is necessary to reproduce the results of Fig.~1a of the main text, where mode 2 is initially populated with a varying number $n_2^{i}$ of photons.    
We conclude that the time-dependent envelope is almost irrelevant when considering the coupling with vacuum, while it is an essential feature to model the SUPER scheme when both modes are initially occupied.

\bibliography{biblio}